\documentclass[12pt]{article}
\begin{document}
\begin{center}{\bf Nonlocal brackets and integrable string models\\}
\vspace{0.3cm}
V.D. Gershun\footnote{e-mail:gershun@kipt.kharkov.ua}
\vspace{0.3cm}
\\{Institute of Theoretical Physics, NSC Kharkov Institute of Physics and 
Technology, Kharkov, UA}\\
\vspace{0.3cm}\end{center}

\begin{abstract}{The closed string model in the background gravity field
is considered as a bi-Hamiltonian system in assumption that string model
is the integrable model for particular kind of the background fields.
 The dual nonlocal Poisson brackets (PB), depending of the background fields
and of their derivatives, are obtained. The integrability condition is
formulated as the compatibility of the bi-Hamiltonity condition and the Jacobi
identity of the dual Poisson bracket. It is shown that the dual brackets and
dual Hamiltonians can be obtained from the canonical PB and from the initial
Hamiltonian by imposing the second kind constraints on the initial
dynamical system, on the closed string model in the constant background
fields, as example. Two types of the nonlocal brackets are introduced. Constant
curvature and time-dependent metrics are considered, as examples. It is shown,
that the Jacobi identities for the nonlocal brackets have particular solution
for the space-time coordinates, as matrix representation of the simple Lie
group.}\end{abstract}
{\bf 1. Introduction.}\\
 The bi-Hamiltonian approach \cite{Gershun:Fadeev,Gershun:Mitropolski,
Gershun:Okubo} to the integrable systems
was initiated by Magri \cite{Gershun:Magri}
for the investigation of the integrability of the KdV equation.\\
{\bf Definition 1.} 
A finite dimensional dynamical system
with $2N$ degrees of freedom $x^{a}, a=1,...2N$ is integrable, if it is
described by the set of the $n$ integrals of motion $F_{1},...,F_{n}$ in
involution under some Poisson bracket (PB)
\[\{F_{i}, F_{k}\}_{PB} = 0.\]
The dynamical system is completely solvable, if $n=N$. Any of the integral
of motion (or any linear combination of them) can be considered as the
Hamiltonian $H_{k} = F_{k}$ .\\
{\bf Definition 2.}
The bi-Hamiltonity condition has following
form:
\begin{equation}\label{Gershun:eq1}\dot x^{a}= \frac{dx^{a}}{dt}=
\{x^{a},H_{1}\}_{1}=...= \{x^{a},H_{N}\}_{N}.\end{equation}
 The hierarchy of new PB is arose in this
connection:
\[{\{ ,\}_{1}, \{ ,\}_{2},..\{ ,\}_{N}}.\]
 The hierarchy of new dynamical systems arises under the new time
coordinates $t_{k}$.
\begin{equation}\label{Gershun:eq2}\frac{dx^{a}}{dt_{n+k}}=
\{x^{a},H_{n}\}_{k+1}=
\{x^{a},H_{k}\}_{n+1}.\end{equation}
The new equations of motion describe the new dynamical
systems, which are dual to the original system, with the dual set of the
integrals of motion. The dual set of the integrals of motion can be
obtained from original one by the mirror transformations and by 
contraction of the integrals of motion algebra. The contraction of the
integral of motion algebra means that the dynamical system belongs to
the orbits of corresponding generators and describes the invariant
subspace. The set of the commuting integrals of motion belongs to Cartan
subalgebra of this algebra. KdV equation is one of the most interesting
examples of the infinite dimensional integrable mechanical systems with
soliton solutions. We consider the dynamical systems with constraints.
 In this case, first kind constraints are generators of the gauge
transformations and they are integrals of motion. First kind constrains
$F_{k}(x^{a})\approx 0$, $k=1,2...$ form the algebra of constraints under
some PB.
\[\{F_{i}, F_{k}\}_{PB}= C_{ik}^{l}F_{l}\approx 0.\]
 The structure functions $C^{l}_{ik}$ may be functions of the
phase space coordinates in general case. The second kind constraints
$f_{k}(x^{a}) \approx 0$ are the representations of the first kind
constraints algebra. The second kind constraints is defined by the
condition
\[\{f_{i}, f_{k}\} = C_{ik}\ne 0.\]
The reversible matrix $C_{ik}$
is not constraint and also it is a function of phase space coordinates. The
second kind constraints take part in deformation of the $\{ , \}_{PB}$ to
the Dirac bracket $\{ , \}_{D}$. As rule, such deformation leads to 
nonlinear and to nonlocal brackets.
 The bi-Hamiltonity condition leads to the dual PB that are nonlinear and
nonlocal brackets as a rule. We suppose, that the dual brackets can be
obtained from the initial canonical bracket under the imposition of the
second kind constraints. We have applied \cite{Gershun:Ger5,Gershun:Ger6,
Gershun:Ger7}, \cite{Gershun:Ger8,Gershun:Ger9}
bi-Hamiltonian approach to the investigation of the integrability of the
closed string model in the arbitrary background gravity field and
antisymmetric B-field. The bi-Hamiltonity condition and the Jacobi
identities for the dual brackets were considered as the integrability
condition for a closed string model. They led to some restrictions on the
background fields. Dubrovin and Novikov have considered the local dual
PB of the similar type \cite{Gershun:Dubrovin} in the
application to the Hamiltonian hydrodynamical models.
The PB of the hydrodynamical type for the phase coordinate functions
$u^{i}(x,t)$ is defined by the formula

\[\{u^{i}(x),u^{k}(y)\}=g^{ik}(u(x))\frac{\partial}{\partial x}\delta (x-y)
-g^{ij}\Gamma^{k}_{jl}(u(x)){u^{l}_x}\delta (x-y).\]

There $g_{ik}(u)$, $\Gamma^{k}_{jl}(u)$ are the arbitrary functions of the 
phase
space coordinates and $u_{x}={\partial}_{x}u$. The Jacobi identity is satisfied
if $g_{ik}$ is the Riemann metric without torsion, the curvature tensor is
equal to zero. The metric tensor is constant, locally.
Mokhov and Ferapontov introduced the nonlocal PB
\cite{Gershun:Mokhov,Gershun:Ferapontov}.
The Ferapontov nonlocal PB is:
\[\{u^{i}(x),u^{k}(y)\}=g^{ik}(u)\frac{\partial}{\partial x}
\delta (x-y)-g^{ij}\Gamma^{k}_{jl}u^{l}_{x}\delta (x-y)+
\omega^{(s)i}_{k}(u(x))u^{j}_{x}\nu (x-y)\omega^{(s)k}_{d}(u(y))u^{d}_{y},
\nonumber\\ \]
where $\nu (x-y)=sgn (x-y)=(\frac{d}{dx})^{-1}\delta (x-y).$
  This PB was used for description of the Hamiltonian
system of the hydrodynamical type. There are systems with functionals of
the hydrodynamical type. The density of this functionals does not depend
on the derivatives $u^{k}_{x}, u^{k}_{xx},... $ and Hamiltonian is also
functional of the hydrodynamical type. On the contrary to these models, the
functionals of the closed string model is depended of the derivatives of
the string coordinates.
The plan of the paper is the following. In the second section we considered
closed string model in the arbitrary background gravity field. We suppose
that this model is an integrable model for some configurations of the 
background fields. The bi-Hamiltonity condition and the Jacobi identities for 
the dual PB must result in to the integrability condition, which  
restrict the possible configurations of the background fields. As examples we
considered constant curvature space and time-dependent metric space.
In the third section we considered closed string model in the constant 
background gravity field\\
{\bf 2. Closed string in the background fields.}\\
 The string model in the background gravity field is described by the
system of the equations:

\[\ddot x^{a}-x^{''a}+\Gamma^{a}_{bc}(x)(\dot x^{b}\dot
x^{c}-x^{'b}x^{'c})=0,\,\, g_{ab}(x)(\dot x^{a}\dot
x^{b}+x^{'a}x^{'b})=0,\,\, g_{ab}(x)\dot x^{a}x^{'b}=0,\nonumber\\ \]

where $\dot x^{a}=\frac{dx^{a}}{d \tau}$, $x^{'a}=\frac{dx^{a}}{d \sigma}$.
 We will
consider the Hamiltonian formalism. The closed string in the background
gravity field is described by first kind constraints in the Hamiltonian
formalism:
\begin{equation}\label{Gershun:eq3}h_{1}=\frac{1}{2}g^{ab}(x)p_{a}p_{b}
+\frac{1}{2}g_{ab}(x)x^{'a}x^{'b}\approx 0,\,\,
h_{2}=p_{a}x^{'a}\approx 0,\end{equation}
where $a,b =0,1,...D-1$, $x^{a}(\tau,\sigma), p_{a}(\tau,\sigma)$ are the
periodical functions on $\sigma$ with the period on $\pi$. The original PB
are the symplectic PB
\[\{x^{a}(\sigma),p_{b}({\sigma}')\}_{1}=\delta _{b}^{a}
\delta (\sigma -{\sigma}'),
\{x^{a}(\sigma),x^{b}({\sigma}'
\}_{1}=\{p_{a}(\sigma),p_{b}({\sigma}') \}_{1}=0.\] The
Hamiltonian equations of motion of the closed string, in the arbitrary
background gravity field under the Hamiltonian
$H_{1}=\int\limits_{0}^{\pi}h_{1}d\sigma$ and PB $\{,\}_{1}$, are
\[\dot x^{a}= g^{ab}p_{b},\,\,\, \dot p_{a}= g_{ab}x^{''b}-
\frac{1}{2}\frac{\partial g^{bc}}{\partial x^{a}}p_{b}p_{c}
-\frac{1}{2}\frac{\partial g_{bc}}{\partial x^{a}}+\frac{\partial
g_{ac}}{\partial x^{b}}.\] The dual PB are obtained from
the bi-Hamiltonity condition
\[\dot
x^{a}=\{x^{a},\int\limits_{0}^{\pi}h_{1}({\sigma}')d{\sigma}' \}_{1}=
\{x^{a},\int\limits_{0}^{\pi}h_{2}({\sigma}')d{\sigma}' \}_{2},\]
\begin{equation}\label{Gershun:eq4}
\dot p_{a}=\{p_{a},\int\limits_{0}^{\pi}h_{1}({\sigma}')d{\sigma}'\}_{1}=
\{p_{a},\int\limits_{0}^{\pi}h_{2}({\sigma}')d{\sigma}'\}_{2}.\end{equation}
They have the following form:\\
{\bf Proposition 1.}\\
\[\{A(\sigma ),B({\sigma}' )\}_{2}= 
\frac{\partial A}{\partial x^{a}}\frac{\partial B}{\partial x^{b}}[[\omega^{ab}
(\sigma )+\omega^{ab}({\sigma}' )]\nu ({\sigma}' -\sigma )+ [\Phi^{ab}(\sigma
)+ \Phi^{ab}({\sigma}' )]\frac{\partial}{\partial {\sigma}' }\delta ({\sigma}'
 -\sigma )\]
\[+[\Omega^{ab}(\sigma )+\Omega^{ab}({\sigma}' )]\delta ({\sigma}' -\sigma )]+
\frac{\partial A}{\partial p_{a}}\frac{\partial B}{\partial p_{b}}[[\omega_{ab}
(\sigma )+\omega_{ab}({\sigma}' )]\nu ({\sigma}' -\sigma )+\] 
\[+[\Phi_{ab}(\sigma )+\Phi_{ab}({\sigma}' )]\frac{\partial}
{\partial {\sigma}' }
\delta ({\sigma}' -\sigma )+ [\Omega_{ab}(\sigma )+\Omega_{ab}({\sigma}' )]
\delta ({\sigma}' -\sigma )]+\]
\[ +[\frac{\partial A}{\partial x^{a}}\frac{\partial B}{\partial
p_{b}}+\frac{\partial A}{\partial p_{b}}\frac{\partial B}{\partial
x^{a}}][[\omega_{b}^{a}(\sigma )+\omega_{b}^{a}({\sigma}' )]\nu ({\sigma}'
-\sigma )+ [\Phi_{b}^{a}(\sigma )+\Phi_{b}^{a}({\sigma}'
)]\frac{\partial}{\partial {\sigma}' }\delta ({\sigma}' -\sigma )]\]
\[ + [\frac{\partial A}{\partial x^{a}}\frac{\partial B}{\partial
p_{b}}-\frac{\partial A}{\partial p_{b}}\frac{\partial B}{\partial x^{a}}]
[\Omega _{b}^{a}(\sigma)+ \Omega_{b}^{a}({\sigma}' )]\delta ({\sigma}' -
\sigma )\]
The arbitrary
functions $A, B, \omega, \Phi, \Omega$ are the functions of the $x^{a}(\sigma
), p_{a}(\sigma )$.  The functions $\omega^{ab}, \omega_{ab}$,
$\Phi^{ab},\Phi_{ab}$ are the symmetric functions on $a, b$ and
$\Omega^{ab}, \Omega_{ab}$ are the antisymmetric functions to satisfy the
condition $\{A, B\}_{2}=-\{B,A\}_{2}$.
 The equations of motion under the
Hamiltonian $H_{2}=\int\limits_{0}^{\pi} h_{2}({\sigma}')d{\sigma}'$ and PB
$\{, \}_{2}$ are
\[ \dot
x^{a}=-\omega_{b}^{a}x^{b}+2\omega^{ab}p_{b}+2\Phi^{ab}p_{b}^{''}
-2\Phi_{b}^{a}x^{''b}+2\Omega_{b}^{a}x^{'b}-2\Omega^{ab}p_{b}^{'}+ \]
\[+ \int\limits_{0}^{\pi}d{\sigma}' [\omega_{b}^{a}x^{'a}+\frac{d
\omega^{ab}}{d{\sigma}'}p_{b}]\nu ({\sigma}' -\sigma )
+\frac{d \Phi^{ab}}{d \sigma}p_{b}^{'}
-\frac{d \Phi_{b}^{a}}{d \sigma}x^{'b},\]
\[ \dot
p_{a}=-\omega_{ab}x^{b}-2\Phi_{ab}x^{''b}+2\Omega_{ab}x^{'b}+
2\omega_{a}^{b}p_{b}+2\Phi_{a}^{b}p_{b}^{''}+2\Omega_{a}^{b}p_{b}^{'}+\]
\[+ \int\limits_{0}^{\pi}d{\sigma}' [\omega_{ab}x^{'b}+\frac{d
\omega^{b}_{a}}{d{\sigma}'}p_{b}]\nu ({\sigma}' -\sigma ) -\frac{d \Phi_{ab}}{d
\sigma}x^{'b}+ \frac{d \Phi_{a}^{b}}{d \sigma}p_{b}^{'}.\]
 The bi-Hamiltonity condition (\ref{Gershun:eq4}) is led to the two constraints
\[
-\omega_{b}^{a}x^{b}+2\omega^{ab}p_{b}+2\Phi^{ab}p_{b}^{''}
-2\Phi_{b}^{a}x^{''b}+2\Omega_{b}^{a}x^{'b}-2\Omega^{ab}p_{b}^{'}+ \]
\[\int\limits_{0}^{\pi}d{\sigma}' [\omega_{b}^{a}x^{'a}+\frac{d
\omega^{ab}}{d {\sigma}'}p_{b}]\nu ({\sigma}' -\sigma )
+\frac{d \Phi^{ab}}{d \sigma}p_{b}^{'}-
\frac{d \Phi_{b}^{a}}{d \sigma}x^{'b}=g^{ab}p_{b},\]

\[-\omega_{ab}x^{b}-2\Phi_{ab}x^{''b}+2\Omega_{ab}x^{'b}+
2\omega_{a}^{b}p_{b}+2\Phi_{a}^{b}p_{b}^{''}+2\Omega_{a}^{b}p_{b}^{'}+\]
\[ + \int\limits_{0}^{\pi}d{\sigma}' [\omega_{ab}x^{'b}+
\frac{d \omega^{b}_{a}}{d{\sigma}'}p_{b}]\nu
({\sigma}' -\sigma ) -\frac{d \Phi_{ab}}{d \sigma}x^{'b}+
\frac{d\Phi_{a}^{b}}{d \sigma}p_{b}^{'}= \]
\[+g_{ab}x^{''b}- \frac{1}{2}\frac{\partial
g^{bc}}{\partial x^{a}}p_{b}p_{c} -\frac{1}{2}\frac{\partial g_{bc}}
{\partial x^{a}}x^{'b}x^{'c}+
\frac{\partial g_{ac}}{\partial x^{b}}x^{'b}x^{'c}.\]

In really, there is the list of the constraints depending on the possible
choice of the unknown functions $\omega, \Omega, \Phi$.
 In the general case, there are both 
the first kind constraints and the second kind constraints. Also it is
possible to solve the constraints equations as the equations for the
definition of the functions $\omega$,$\Phi$,$\Omega$. We considered the
latter possibility and we obtained the following consistent solution of the
bi-Hamiltonity condition:
\begin{eqnarray}\Phi^{ab}=0,\,\,\Omega^{ab}=0,\,\,\Phi_{b}^{a}=0,\,\,
\Omega ^{a}_{b}=0,\,\,
\frac{\partial\omega^{ab}}{\partial x^{c}}x^{c}+2\omega^{ab}=g^{ab},\nonumber\\
\omega_{ab}=\frac{1}{2}\frac{\partial^{2} \omega^{cd}}{\partial x^{a}\partial
x^{b}}p_{c}p_{d},\,\,\omega_{b}^{a}=-\frac{\partial \omega^{ac}}{\partial
x^{b}}p_{c},\nonumber \\
\Phi_{ab}=-\frac{1}{2}g_{ab},
\Omega_{ab}=\frac{1}{2}(\frac{\partial \Phi_{bc}}{\partial x^{a}}-
\frac{\partial
\Phi_{ac}}{\partial x^{b}})x^{'c},\,\, \frac{\partial\omega^{ab}}
{\partial p_{c}}=0.
\nonumber\end{eqnarray}\\
{\bf Remark 1.}
In distinct from the PB of the
hydrodynamical type, we need to introduce the separate PB for the
coordinates of the Minkowski space and for the momenta because, the
gravity field is not depend of the momenta. Although, this difference is
vanished under the such constraint as $f(x^{a},p_{a})\approx 0$.\\
Consequently, the dual PB for the phase
space coordinates are
\[\{x^{a}(\sigma
),x^{b}({\sigma}')\}_{2}= [\omega^{ab}(\sigma )+\omega^{ab}({\sigma}' )]\nu
({\sigma}' - \sigma ),\]
\[ \{p_{a}(\sigma ),p_{b}({\sigma}' )\}_{2}=
[\frac{\partial^{2} \omega_{cd}(\sigma )}{\partial x^{a}\partial
x^{b}}p_{c}p_{d}+\frac{\partial^{2} \omega_{cd}({\sigma}' )}{\partial
x^{a}\partial x^{b}}p_{c}p_{d}]\nu ({\sigma}' - \sigma )- \]
\[-\frac{1}{2}[g_{ab}(\sigma)+g_{ab}({\sigma}')]\frac{\partial}{\partial
{\sigma}' }\delta ({\sigma}' - \sigma )+
[\frac{\partial g_{ac}}{\partial
x^{b}}-\frac{\partial g_{bc}}{\partial x^{a}}]x^{'c}(\sigma)\delta
({\sigma}' - \sigma)\]
\[ \{x^{a}(\sigma ),p_{b}({\sigma}' )\}_{2}= -
[\frac{\partial \omega^{ac}(\sigma )}{\partial
x^{b}}p_{c}+\frac{\partial \omega^{ac}({\sigma}')}{\partial x^{b}}p_{c}
]\nu ({\sigma}' - \sigma ),\]
\begin{equation}\label{Gershun:eq5}\{p_{a}(\sigma ),x^{b}({\sigma}'
)\}_{2}= - [\frac{\partial \omega^{bc}(\sigma)}{\partial
x^{a}}p_{c}+\frac{\partial \omega^{bc}({\sigma}' )}{\partial x^{c}
}p_{c}]\nu ({\sigma}' -\sigma ).\end{equation}
 The function
$\omega^{ab}(x)$ is satisfied on the equation:
\begin{equation}\label{Gershun:eq6}\frac{\partial \omega^{ab}}{\partial
x^{c}}x^{c}+2\omega^{ab} =g^{ab}.\end{equation} 
The Jacobi identities for the PB $\{,\}_{2}$ are led to the nonlocal
consistence conditions on the unknown function $\omega^{ab}(\sigma)$.
We can calculate unknown metric tensor $g^{ab}(\sigma)$ by substitution
of the solution of the consistence condition for $\omega^{ab}$ to the
equation (\ref{Gershun:eq6}).
  The Jacobi identity
\begin{equation}\label{Gershun:eq7}
\{x^{a}(\sigma),x^{b}({\sigma}')\}x^{c}({\sigma}'')\}_{J}\equiv\end{equation}
\[\{x^{a}(\sigma),x^{b}({\sigma}')\}x^{c}({\sigma}'')\}
+\{x^{c}({\sigma}''),x^{a}(\sigma)\}x^{b}({\sigma}')\}+\{x^{b}({\sigma}'),
x^{c}({\sigma}'')\}x^{a}(\sigma\})=0\]
is led to the following nonlocal analogy of the WDVV 
\cite{Gershun:Wit,Gershun:Verl} consistence condition:
\[[\frac{\partial \omega ^{ab}(\sigma)}{\partial
x^{d}}[\omega^{dc}(\sigma)+\omega^{dc}({\sigma}'')]- \frac{\partial
\omega^{ac}(\sigma)}{\partial x^{d}}[\omega
^{db}(\sigma)+\omega^{db}({\sigma}')]]\nu ({\sigma}'-\sigma)\nu ({\sigma}''-
\sigma)+\]
\[ [\frac{\partial \omega ^{cb}({\sigma}')}{\partial
x^{d}}[\omega^{da}({\sigma}')+\omega^{da}(\sigma)]-
 \frac{\partial
\omega^{ab}({\sigma}')}{\partial x^{d}}[\omega
^{dc}({\sigma}')+\omega^{dc}({\sigma}'')]]\nu (\sigma-{\sigma}')\nu ({\sigma}''
- {\sigma}')+\nonumber\]
\begin{equation}\label{Gershun:eq8}[\frac{\partial \omega^{ac}({\sigma}'')}
{\partial x^{d}}
[\omega^{db}({\sigma}'')+\omega^{db}({\sigma}')]-
\frac{\partial \omega^{cb}
({\sigma}'')}{\partial x^{d}}[\omega^{da}({\sigma}'')+ \omega^{da}(\sigma)]]
\nu (\sigma-{\sigma}'')\nu ({\sigma}'-{\sigma}'')=0.\end{equation}

  This equation has the particular solution of the following form:
\[\label{Gershun:eq9}\frac{\partial\omega^{ab}(\sigma)}
{\partial x^{d}}
[\omega^{dc}(\sigma)+\omega^{dc}({\sigma}'')]-\frac{\partial \omega^{ac}
(\sigma)}{\partial x^{d}}[\omega^{db}(\sigma)+\omega^{db}({\sigma}')]=
\nonumber \]
\[ [T^{b},T^{c}]T^{a}]f(\sigma,{\sigma}',{\sigma}'')\nu({\sigma}''-
\sigma)\nu ({\sigma}'-\sigma),\]
where $T^{a}, a = 0,1,...D-1$ is the matrix representation of the simple Lie 
algebra and $f(\sigma,{\sigma}',{\sigma}'')$ is arbitrary function.
The Jacobi identity is satisfied on the Jacobi identity of the simple Lie
algebra in this case:
\[([T^{a},T^{b}]T^{c}]+[T^{c},T^{a}]T^{b}]+
[T^{b},T^{c}]T^{a}])f(\sigma,{\sigma}',{\sigma}'')=0\]
and we used the relation ${\nu}^{2}({\sigma}'-\sigma)=1$.
 The local solution of the Jacobi identities leads to the constant metric
tensor. The rest Jacobi identities are cumbrous and we do not reduce this
expressions here.
 The symmetric factor of $\sigma,{\sigma}'$ of the antisymmetric functions
$\nu({\sigma}'-\sigma)$,$\frac{\partial}{\partial
\sigma}\delta(\sigma-{\sigma}')$ in the right side of the PB can be both
sum of the functions of $\sigma$ and ${\sigma}'$, and production of them.
 Last possibility can be used in the vielbein formalism.\\
{\bf Proposition 2.}
The bi-Hamiltonity condition can be solved in the terms PB $\{,\}_{2}$,
which have the following form:
\[\{x^{a}(\sigma),x^{b}({\sigma}')\}_{2}=
e^{a}_{\mu}(\sigma)e^{b}_{\mu}({\sigma}')\nu ({\sigma}'-\sigma),\]
\[\{x^{a}(\sigma),p_{b}({\sigma}'\}_{2}=-e^{a}_{\mu}
(\sigma)\frac{\partial e^{c}_{\mu}({\sigma}'}{\partial x^{b}}p_{c}({\sigma}')
\nu ({\sigma}'-\sigma),\]
\[\{p_{a}(\sigma),p_{b}({\sigma}')\}_{2}=\frac{\partial
e^{c}_{\mu}(\sigma)}{\partial x^{a}}p_{c}(\sigma)\frac{\partial e^{d}_{\mu}
({\sigma}')}{\partial x^{b}}p_{d}({\sigma}')\nu ({\sigma}'-\sigma)-
e_{a}^{\mu}(\sigma)e_{b}^{\mu}({\sigma}')\frac{\partial }{\partial {\sigma}'}
\delta ({\sigma}'-\sigma)+\]
\begin{equation}\label{Gershun:eq10}+[\frac{\partial e_{a}^{\mu}}
{\partial x^{c}}e_{b}^{\mu}
-\frac{\partial e_{b}^{\mu}}{\partial x^{c}}e_{a}^{\mu}-\frac
{\partial e_{c}^{\mu}}{\partial x^{a}}e_{b}^{\mu}+
\frac{\partial e_{c}^{\mu}}{\partial x^{b}}e_{a}^{\mu}]
x^{'c}(\sigma)\delta ({\sigma}'-\sigma),\end{equation}
where veilbein $e^{a}_{\mu}$ is satisfied on the additional conditions:
\[g^{ab}= \eta ^{\mu\nu}e^{a}_{\mu}e^{b}_{\nu},\,\,
g_{ab}=\eta_{\mu\nu}e_{a}^{\mu}e_{b}^{\nu}\]
and $\eta^{\mu\nu}$ is the metric tensor of the flat space.\\
 The particular solution of the Jacobi identity is
\[\label{Gershun:eq11}\frac{\partial e^{a}_{\mu}(\sigma)}
{\partial x^{d}}e^{b}_{\mu}({\sigma}')
e^{d}_{\nu}(\sigma)e^{c}_{\nu}({\sigma}'')-\frac{\partial e^{a}_{\mu}
(\sigma)}{\partial x^{d}}e^{c}_{mu}({\sigma}'')e^{d}_{\nu}(\sigma)e^{b}_
{\nu}({\sigma}')=\]
\[[T^{b},T^{c}]T^{a}]f(\sigma,{\sigma}',{\sigma}'')\nu ({\sigma}''-\sigma)
\nu ({\sigma}'-\sigma).\]
As example let me consider the the constant curvature space.\\
{\bf Example 1.}
The constant curvature space is
described by the metric tensor $g_{ab}(x(\sigma))$ and by it inverse tensor
$g^{-1}_{ab}$:
\[g_{ab}=\eta_{ab}+\frac{kx_{a}x_{b}}{1-kx^{2}}, \,\,
g^{ab}\equiv g^{-1}_{ab}=\eta_{ab}-kx_{a}x_{b}.\]\\
{\bf Proposition 3.}
Dual (PB)$\{,\}_{2}$ are:
\[\{x_{a}(\sigma),x_{b}({\sigma}')\}=[\eta _{ab}-kx_{a}
(\sigma)x_{b}({\sigma}')]
\nu ({\sigma}'-\sigma),\]
\[\{x_{a}(\sigma),p_{b}({\sigma}')\}=
kx_{a}(\sigma)p_{b}({\sigma}')\nu ({\sigma}'-\sigma),\]
\[\{p_{a}(\sigma),p_{b}({\sigma}')\}=-kp_{a}(\sigma)p_{b}
({\sigma}')\nu ({\sigma}'- \sigma)\]
\begin{equation}\label{Gershun:eq12}-\frac{1}{2}[2\eta _{ab}+
\frac{kx_{a}x_{b}}{1-kx^{2}}(\sigma)+
\frac{kx_{a}x_{b}}{1-kx^{2}}({\sigma}')]\frac{\partial}{\partial {\sigma}'}
\delta ({\sigma}'-\sigma)+
\frac{x_{a}x'_{b}-
x_{b}x'_{a}}{2(1-kx^{2})}\delta ({\sigma}'-\sigma).\end{equation}

The Jacobi identity (\ref{Gershun:eq7}) is led to the equation
\[[\eta_{ab}x_{c}({\sigma}'')-\eta_{ac}x_{b}({\sigma}')]
\nu ({\sigma}'-\sigma)\nu (\sigma-{\sigma}'')+[\eta_{bc}x_{a}(\sigma)-
\eta_{ba}x_{c}({\sigma}'')]
\nu (\sigma-{\sigma}')\nu ({\sigma}'-{\sigma}'')+\nonumber \]
\[[\eta_{ca}x_{b}({\sigma}')
-\eta_{cb}x_{a}(\sigma)]
\nu ({\sigma}'-{\sigma}'')\nu ({\sigma}''-\sigma)=0.\nonumber \]
The particular solution of this equation is:
\begin{equation}\label{Gershun:eq13}\eta_{ab}x_{c}({\sigma}'')-\eta_{ac}x_{b}
({\sigma}')=
[T_{b},T_{c}]T_{a}]f(\sigma,{\sigma}',{\sigma}'')\nu
({\sigma}''-\sigma)\nu ({\sigma}'-\sigma).\end{equation}
 Consequently, the space-time coordinate $x_{a}(\sigma)$ is the matrix
representation of the simple Lie algebra.
  The Jacobi identity
 $\{x_{a}(\sigma),x_{b}({\sigma}')\}p_{c}({\sigma}'')\}_{J}$ is led to
the equation
\begin{equation}\label{Gershun:eq14}k\eta_{ab}p_{c}({\sigma}'')\nu
({\sigma}'-\sigma)[\nu ({\sigma}''-\sigma)+\nu ({\sigma}''-{\sigma}')]=0.
\end{equation}
 These results can be obtained from the veilbein formalism under the following
ansatz for the veilbein of the constant curvature space:
\[e^{a(s)}_{\mu}=n_{\mu}(m^{(s)}_{1}n^{a}+
\sqrt{-k}m^{(s)}_{2}x^{a}),
e_{a}^{\mu (s)}=n^{\mu}g_{ab}(m^{(s)}_{1}n^{b}+\sqrt{-k}m^{(s)}_{2}x^{b}),
\]
where $n^{2}_{\mu}=1,\,\,m^{(s)}_{1}m^{(s)}_{1}=1,\,\,
m^{(s)}_{2}m^{(s)}_{2}=1,\,\,m^{(s)}_{1}m^{(s)}_{2}=0,\,\,n^{a}n^{b}=
\delta^{ab}$ and $(s)$ is number of the solution of the equations
\[e^{a}_{\mu}e^{b}_{\mu}=g^{ab},\,\,e_{a}^{\mu}e_{b}^{\mu}=g_{ab},\,\,
e^{a}_{\mu}e^{\mu}_{b}=\delta ^{a}_{b}.\]
  The following example is time-dependent metric space.\\
{\bf Example 2.}
The time-dependent metric in the light-cone variables has form:
\begin{equation}ds^{2}=g_{ik}(x^{+})dx^{i}dx^{k}+g_{++}(x^{+})dx^{+}dx^{+}
+ 2g_{+-}dx^{+}dx^{-}.\nonumber\end{equation}
We are used Poisson brackets (\ref{Gershun:eq5}) for the space coordinates
$x^{a}=
\{x^{i},x^{+},x^{-}\},\,i=1,2...D-2.$ We introduced the light-cone gauge as
two first kind constraints:
\[F_{1}(\sigma)=x'^{+}\approx 0,\,\,F_{2}(\sigma)=p'_{-}\approx 0,\]
and we imposed them on the equations of motion and on the Jacobi identities.
The Jacobi identities are reduced to the simple equation
\[\frac{\partial \omega^{ab}}{\partial x^{+}}\omega^{+c}-\frac{\partial
\omega^{ac}}{\partial x^{+}}\omega^{+b}=0.\]
We obtained following result from this equation and additional condition
(\ref {Gershun:eq6}): there is constant background gravity field only for the
non-degenerate metric.\\
{\bf 3. Constant background fields.}\\
 In this section we are supplemented the bi-Hamiltonity condition
(\ref{Gershun:eq4})
by the mirror transformations of the integrals of motion.
\[\dot x^{a}=\{x^{a},\int\limits_{0}^{\pi}h_{1}d\sigma\}_{1}=
\{x^{a},\int\limits_{0}^{\pi}{\pm h_{2}}d{\sigma}'\}_{\pm 2}.\]
 The dual PB are
\[\{x^{a}(\sigma ),x^{b}({\sigma}')\}_{\pm 2}=
\pm g^{ab}\nu ({\sigma}' -\sigma ),\,\, \{x^{a}(\sigma),p_{b}
({\sigma}' )\}_{\pm 2}=0,\]
\[\{p_{a}(\sigma ),p_{b}({\sigma}' )\}_{\pm 2}
=\mp g_{ab}\frac{\partial}{\partial {\sigma}' }\delta ({\sigma}'
-\sigma ).\]
 The dual dynamical system
\[\dot x^{a}=\{x^{a}, \pm H_{2}\}_{1}=\{x^{a}, H_{1}\}_{\pm 2}.\]
is the left(right) chiral string
\[\dot x^{a}=\pm x^{'a},\,\, \dot p_{a}=\pm p_{a}^{'}.\]
 In the terms of the Virasoro operators
\[
L_{k}=\frac{1}{4\pi}\int\limits_{0}^{\pi}(h_{1}+h_{2})e^{ik\sigma}d\sigma ,\,\,
\bar L_{k}=\frac{1}{4\pi}\int\limits_{0}^{\pi}(h_{1}-h_{2})e^{ik\sigma}d\sigma
,\]
the first kind constraints form the $Vir\oplus Vir$ algebra under the PB
$\{, \}_{1}$.
\[\{L_{n},L_{m}\}_{1}=-i(n-m)L_{n+m},\,\,
\{\bar L_{n},\bar L_{m}\}_{1}=-i(n-m)\bar L_{n+m},\,\,
\{L_{n},\bar L_{m}\}_{1}=0.\]
The dual set of the integrals of motion is obtained from initial it by the
mirror transformations
\[H_{1}\to \pm H_{2},\,\,L_{0}\to \pm  L_{o},\,\,
\bar L_{0}\to \mp \bar L_{0},\,\,\tau \to \sigma. \]
and by the contraction of the first kind constraints algebra $L_{n}=0$, or
$\bar L_{n}=0$, $n\ne 0$.
 Another way to obtain the dual brackets is the imposition of the second
kind constraints on the initial dynamical system, by such manner, that
$F_{i}=F_{k}$ for $i\ne k, i, k = 1,2,...$ on the constraints surface
$f(x^{a},p_{a})=0$.

The constraints $f^{(-)}_{a}(x,p)=p_{a}-g_{ab}x'^{b}\approx 0$ or
$f^{(+)}_{a}=p_{a}+g_{ab}x'^{b}\approx 0$ (do not simultaneously) are the
second kind constraints.  \[\{f^{(\pm)}_{a}(\sigma),
f^{(\pm)}_{b}({\sigma}')\}_{1}=C^{(\pm)}_{ab}(\sigma -{\sigma}')=
\pm 2g_{ab}\frac{\partial}{\partial {\sigma}'}\delta ({\sigma}' -\sigma).\]
 The inverse matrix $(C^{(\pm)})^{-1}$ has following form
$C^{(\pm)ab}(\sigma -{\sigma}')=\pm\frac{1}{2}g^{ab}\nu({\sigma}' -\sigma).$
 There is only one set
of the constraints, because consistency condition 
\[\{f^{(\pm)}(\sigma),
H_{1}\}_{1}= f^{'(\pm)}(\sigma)\approx 0,\,\,...\,\,,
\{f^{(\pm)(n)}(\sigma), H_{1}\}_{1}= f^{(\pm)(n+1)}(\sigma)
\approx 0.\nonumber\] 
is not produce the new sets of constraints.
 By using the
standard definition of the Dirac bracket, we are obtained following Dirac
brackets for the phase space coordinates.  \[\{x^{a}(\sigma),
x^{b}({\sigma}')\}_{D}=\pm\frac{1}{2}g^{ab}\nu({\sigma}' -\sigma),
\{p_{a}(\sigma),p_{b}({\sigma}')\}_{D}=\mp\frac{1}{2}g_{ab}\frac{\partial}
{{\sigma}'}\delta({\sigma}'-\sigma),\]
\[\{x^{a}(\sigma),p_{b}({\sigma}')\}_{D}=\frac{1}{2}\delta^{a}_{b}
\delta({\sigma}' -\sigma).\]
{equation}
 The equations of motion under the Hamiltonians $H_{1}=h_{1},H_{2}= h_{2}$ and
Dirac bracket \[\dot x^{a}=\{x^{a}, H_{1}\}_{D}=\{x^{a},
H_{2}\}_{D}=g^{ab}p_{b}=\pm x'^{a},\]
\[\dot p_{a}=\{p_{a},H_{1}\}_{D}=\{p_{a}, H_{2}\}_{D}=g_{ab}x'^{b}=
\pm p'_{a}.\] are
coincide on the constraints surface. The dual brackets $\{, \}_{\pm 2}$
are coincide with the Dirac brackets also. The contraction of the algebra
of the first kind constraints means that the integrals of motion
$H_{1}=H_{2}$ are coincide on the constraints surface too.

\end{document}